\documentclass{article}
\usepackage{spconf}

\usepackage[ left=1in, top=1in, right=1in, bottom=1in]{geometry}
\usepackage{graphicx,bm,colonequals,amsmath,amssymb,url,xcolor,bbm}
\usepackage{array,tabularx,multirow}
\usepackage{enumitem,MnSymbol}
\usepackage[font={footnotesize}]{subcaption}
\usepackage[utf8]{inputenc}
\usepackage{soul} 
\usepackage{placeins} 
\usepackage[normalem]{ulem}
\usepackage{float}
\usepackage{algorithm, algorithmic}

\graphicspath{{plots/}}

\usepackage{hyperref}
\usepackage[authoryear]{natbib}

\usepackage{mathtools}
\mathtoolsset{showonlyrefs} 

\bibpunct[, ]{(}{)}{;}{a}{,}{,}
   
\usepackage{amsthm}
\usepackage{array,framed}

\makeatletter
\renewcommand{\paragraph}{%
  \@startsection{paragraph}{4}%
  {\z@}{2ex \@plus 1ex \@minus .2ex}{-1em}%
  {\normalfont\normalsize\bfseries}%
}
\makeatother

\usepackage{etoolbox}
\makeatletter
\preto{\@verbatim}{\topsep=0pt \partopsep=0pt }
\makeatother

\DeclareMathAlphabet\mathbfcal{OMS}{cmsy}{b}{n}

\newcommand{\bx}{\mathbf{x}}
\newcommand{\by}{\mathbf{y}}

\newcommand{\bZ}{\mathbf{Z}}

\newcommand{\bY}{\mathbf{Y}}

\newcommand{\bL}{\mathbf{L}}

\newcommand{\bU}{\mathbf{U}}

\newcommand{\bK}{\mathbf{K}}

\newcommand{\bfmu}{\bm{\mu}}

\newcommand{\bfSigma}{\bm{\Sigma}}

\newcommand{\GP}{\mathcal{GP}}

\newcommand{\order}{\mathcal{O}}
\newcommand{\normal}{\mathcal{N}}

\DeclareMathOperator*{\argmin}{arg\,min}

\newcommand{\dens}{p}
\newcommand{\adens}{\widehat p}

\newcommand{\domain}{\mathcal{X}}

\usepackage{comment}
\begin{document}
\title{Scalable Model-Based Gaussian Process Clustering}
%
\name{Anirban Chakraborty$^{\star}$, Abhisek Chakraborty$^{\star}$\thanks{Both the authors contributed equally . There was no external on internal funding for this work..}}
\address{Department of Statistics, Texas A$\&$M University\\ 3143, College Station, TX 77843}
%
%
%
%
\maketitle
\begin{abstract}
Gaussian process is an indispensable tool in clustering functional data, owing to it's flexibility and inherent uncertainty quantification. However,  when the functional data is observed over a large grid (say, of length $p$), Gaussian process clustering quickly renders itself infeasible, incurring $O(p^2)$ space complexity and  $O(p^3)$ time complexity per iteration; and thus prohibiting it's natural adaptation to large environmental applications \citep{tgpm, stgpm}. To ensure scalability of Gaussian process clustering in such applications, we propose to embed the popular Vecchia approximation \citep{Vecchia1988} for Gaussian processes at the heart of the  clustering task, provide crucial theoretical insights towards algorithmic design, and finally develop a computationally efficient expectation maximization (EM) algorithm. Empirical evidence of the utility of our proposal is provided via simulations and analysis of polar temperature anomaly  (\href{https://www.ncei.noaa.gov/access/monitoring/climate-at-a-glance/global/time-series}{noaa.gov}) data-sets. 
\end{abstract}
\begin{keywords}
Functional Data, Gaussian Process Mixtures, Kullback-Leibler Projection, Vecchia Approximation, temperature Anomaly.
\end{keywords}
\section{Introduction}
\label{sec:intro}
Functional data clustering \citep{fda} aims to discern distinct patterns in underlying continuous functions based on observed discrete measurements over a grid. Model-based methods for clustering functional data \citep{functionalsubspace2011, funsclust2013} are popular in applications in engineering, environmental sciences, social sciences etc; since the approaches allow for assumption of complex yet parsimonious covariance structures, and perform simultaneous dimensionality reduction and clustering. Such methods crucially involve modeling the functional principal components scores \citep{functionalsubspace2011, funsclust2013} or  coefficients of basis functions \citep{f5e8e9a1-2f12-3d9d-bc1f-cdebe487b68a}. 

Recent literature on functional data clustering has gravitated towards adopting Gaussian processes \citep{rasmussen}, especially in the context of environmental applications \citep{tgpm, stgpm}, owing to it's flexibility, interpretabilty, and natural uncertainty quantification. However, naive implementation of Gaussian process clustering \citep{tgpm} incurs $O(p^2)$ space complexity and  $O(p^3)$ time complexity, and becomes infeasible for data sets observed over large grids. To circumnavigate this issue, we appeal to the recent literature on scalable computation involving Gaussian processes.

In spatial statistics, the Vecchia approximation \citep{Vecchia1988} and its extensions \citep{10.1214/19-STS755, 10.1007/s13253-020-00401-7} are popular ordered conditional approximations of Gaussian processes, that imply a valid joint distribution for the data and results in straightforward likelihood-based parameter inference. This allows for proper uncertainty quantification in downstream applications. On the computational front, Vecchia approximation of Gaussian processes leads to inherently parallelizable operations and results in massive computational gains, refer to \ref{sec: vecchiaintro} for further details.  Consequently, Vecchia approximation of Gaussian processes is adopted in a plethora of applications in recent times, e.g, Bayesian optimization \citep{pmlr-v206-jimenez23a}, Computer model emulation \citep{doi:10.1137/20M1352156}, Gaussian process regression, to name a few. 

In this article, to ensure scalability of Gaussian process clustering in large scale applications, we propose to adopt the popular Vecchia approximation for Gaussian processes at every iteration of the clustering, provide crucial theoretical framework for a potential algorithmic design, and finally develop a computationally efficient expectation maximization algorithm. Clustering accuracy and computational gains of the proposed algorithm are delineated through several numerical experiments, and publicly available data set on temperature anomaly in the Earth's geographic north pole.

\section{Proposed Methodology}
\subsection{Gaussian Process (GP) Mixture Models \label{sec:finitegpmixture}}
Let $y(\bx)$ be the observed output at a $d$-dimensional input vector $\bx\in\domain$, and $y(\cdot)\sim \GP(\bfmu,\bK)$ is assumed to be a Gaussian process ($\GP$) \citep{rasmussen} with mean function $\bfmu: \domain \rightarrow \mathbb{R}$ and a positive-definite covariance  function $\bK: \domain \times \domain \rightarrow \mathbb{R}$. Throughout most of this article, for the sake of brevity, we focus on exponential covariance function of the form
$\bK(\bx_i,\bx_j)$ = $\sigma^2 \mbox{exp} \left(-\frac{(x_i - x_j)^2}{l^2}\right)$ 
with range parameter $l$ and scale parameter $\sigma$; however the proposed methodology generalises beyond the specific choice. By definition of $\GP$, the vector $\by = \big(y(\bx_1),\ldots,y(\bx_p)\big)^\top$ of responses at $p$ input values $\{\bx_1,\ldots,\bx_p\}$ follows an $p$-variate Gaussian distribution with covariance matrix
$
\bK = \big(K(\bx_i,\bx_j)\big)_{i,j=1,\ldots,p},
$ 
whose $(i,j)^{\mbox{th}}$ entry describes the covariance between the responses  corresponding to the input values $\bx_i$ and $\bx_j$.

Finite mixtures of such Gaussian processes \citep{tgpm, stgpm} provide a flexible tool to model functional data, and it takes the form $y(\cdot)\sim$ $\sum_{g=0}^G \pi_g \GP (\bfmu_g, \bK_g)$, , where  $\pi_g\geq 0, g=1,\ldots, G;\ \sum_{g=1}^G\pi_g =1$ . Again, by definition of $\GP$, a vector $\by = \big(y(\bx_1),\ldots,y(\bx_p)\big)^\top$ of responses at $p$ input values $\{\bx_1,\ldots,\bx_p\}$ follows a $G$-mixture of $p$-variate Gaussian distributions $y(\cdot)\sim \sum_{g=0}^G \pi_g \normal_p (\bfmu_g, \bK_g)$. Estimation of the  model parameters $\mu_g, \pi_g, l_g, \sigma_g, \forall g = 1, \ldots, G$ is routinely carried out via Expectation-maximization (EM) algorithm \citep{dempster1977em}.

\subsection{Naive EM Algorithm \label{sec:naiveem}}
Suppose, we observe $N$ independent realizations $\bY = (\by_1, \by_2, \ldots, \by_N)$ of GP mixtures, where $\by_i$ is the output corresponding to a $p$-dimensional input grid of a covariate $x\in\mathcal{R}$, i.e., $\by_i(x) = (\by_i(x_1), \by_i(x_2), \ldots, \by_i(x_p))$. To classify the realizations into $G$ (known) separate clusters, we first present a naive EM algorithm  \citep{dempster1977em} and delineate associated computational bottlenecks. Since a $\GP$ realization on a grid follows a multivariate Gaussian distribution, the complete log-likelihood takes the form,
\begin{align}
 \label{eq:completell}
    \mathcal{L}(\theta | \bY)
    &= \sum_{i=1}^N\sum_{g=1}^G1_{\{z_i = g\}} \mbox{log}\left(\pi_g \normal_p (y_i; \bfmu_g, \bK_g)\right)
\end{align}
where $\theta = (\pi_{1:G}, \mu_{1:G}, l_{1:G}, \sigma_{1:G})$ denote the model parameters of interest, and $\bZ = (z_1, \ldots, z_N)$ are the latent clustering indices.

Without loss of generality, we assume $\bfmu_g = 0 \forall g$, and present the EM-updates for $\theta = (\pi_{1:G}, l_{1:G}, \sigma_{1:G})$.
\textbf{(1) E- Step.} The conditional expectation at $s^{th}$ iteration, $Q(\theta | \theta^{s})= \mathbf{E}_{z_1, \ldots z_N \sim p( . |\bY, \theta^{(s)})} \mathcal{L}(\bZ, \theta | \bY) = \sum_{i=1}^N \sum_{g=1}^G\left( W^{(s)}_{ig} \mbox{log}(\pi_g 
\normal(\by_i; 0, \bK_g))\right)$,
where $W^{(s)}_{ig} = \mathbf{P}(z_i = g|\bY, \theta^{(s)}) \propto \pi_g \normal_p (y_i; 0, \bK^{(s)}_g)$.
\textbf{(2) M-step.} Next, we calculate $\theta^{(s+1)}$ = $\underset{\theta}{\mbox{argmax}}\ Q(\theta | \theta^{(s)})$,
via gradient ascent  updates
$\pi^{(s+1)}_g \propto\sum_{i=1}^N W^{(s)}_{ig},\
l^{(s+1)}_g = l^{(s)}_g + \lambda\frac{d}{dl_g}\epsilon^{(s)},\ 
\sigma^{(s+1)}_g = \sigma^{(s)}_g + \lambda\frac{d}{d\sigma_g}\epsilon^{(s)},
$
where $\epsilon^{(s)} = \sum_{i=1}^N\sum_{g=1}^G \frac{W^{(s)}_{ig}}{2}\left(\mbox{log}|\bK^{-1}_g| - \by^\top_i \bK^{-1}_g \by_i\right)$, and $\lambda$ denotes the learning rate. We run the two steps iteratively until convergence. 

Although the above algorithm straightforward to implement, both the \textbf{E} and \textbf{M} steps require inversion of covariance kernels in the normal log-likelihood calculation, which incurs a $\order(p^3)$ time complexity. While spectral decompositions \citep{spectral} can speed up matrix inversion, these methods often fail to work in practice even for matrices of moderate dimensions. Alternatively, to develop our modified EM algorithm, we  utilize the popular Vecchia approximation in evaluating the log-likelihood in \eqref{eq:completell},  which enable us to carry out matrix inversions in quasi-linear time.   

\subsection{Vecchia Approximation for GP \label{sec: vecchiaintro}}
Motivated by the exact decomposition of the joint density $\dens(\by) = \prod_{i=1}^p \dens(y_i|\by_{1:i-1})$ as a product of univariate conditional densities, \citep{Vecchia1988} proposed the approximation
\begin{equation}
    \label{eq:vecchia}
\textstyle \adens(\by) = \prod_{i=1}^p \dens(y_i|\by_{c(i)}),
\end{equation}
where $c(i) \subset \{1,\ldots,i-1\}$ is a conditioning index set of size $|c(i)| = \min(m,i-1)$ for all $i=2,\ldots,n$ (and $c(1) = \emptyset$). Even with relatively small conditioning-set size $m\ll p$, the approximation \eqref{eq:vecchia} with appropriate choice of the $c(i)$ is very accurate due to the screening effect \citep{Stein2011b}. We get back the exact likelihood for $m=p-1$.  The approximation accuracy of the Vecchia approach depends on the choice of ordering of the variables $y_1,\ldots,y_n$ and on the choice of the conditioning sets $c(i)$. A general Vecchia framework \citep{Katzfuss2017a} obtained by varying these choices unifies many popular GP approximations \citep{Quinonero-Candela2005, Snelson2007}.
In practice, high accuracy can be achieved using a maximum-minimum distance (maximin) ordering \citep{Guinness2016a}, that picks the first variable arbitrarily, and then chooses each subsequent variable in the ordering as the one that maximizes the minimum distance to previous variables in the ordering, and the distance between two variables $y_i$ and $y_j$ is  defined as the Euclidean distance $\|\bx_i - \bx_j\|$ between their corresponding inputs.

The choice of the conditioning set $c(i)$ implies a sparsity constraint $\mathcal{S}$ on the covariance structure of $\textstyle \adens(\by)$. \citep{doi:10.1137/20M1336254} showed that, under $\mathcal{S}$, the Vecchia approximation $\textstyle \adens(\by)$ of the true $\GP$ $p(\by)$ is the optimal Kullback-Leibler (KL) projection of  the true $\GP$ $p(\by)$ within the class of $\GP$s that satisfy $S$, i.e,
$
    \adens(\by)  = \argmin_{L\in \mathcal{S}} \mbox{KL}[p(\mathbf{y}) \ ||\ \normal_p(\bfmu, (LL^{T}))^{-1}],
$
where $\mathcal{S} = \{ L\in \mathcal{R}^{p\times p}: L_{ij} \neq 0\implies (i,j)\in S$\}. The implied approximation a grid, $\adens(\by) = \normal_p(\bfmu,\widehat\bK)$ is also multivariate Gaussian, and the Cholesky factor of $\widehat\bK^{-1}$ is highly sparse with fewer than $pm$ off-diagonal nonzero entries \citep{Datta2016,Katzfuss2017a}. Consequently, the $\dens(y_i|\by_{c(i)})$ in \eqref{eq:vecchia} are all Gaussian distributions that can be computed in parallel using standard formulas, each using $\order(m^3)$ operations.

\subsection{Vecchia Approximation for GP Mixtures}
 To improve upon  the EM algorithm in sub-section \ref{sec:naiveem}, we develop a Vecchia approximation framework for Gaussian process mixture models of the form $p^{(\mbox{gpm})}(\mathbf{y}) = \sum_{g=1}^G \pi_g p_g(\mathbf{y})$. Motivated by \ref{eq:vecchia}, for a fixed conditioning set size $m$, we express $p^{(\mbox{gpm})}(\mathbf{y})$ as mixture of products of univariate conditional densities, $\textstyle \adens ^{(\mbox{gpm, vechhia(m)})}(\by) = \sum_{g=1}^G \pi_g[\prod_{i=1}^p \dens_k(y_i|\by_{c(i)})]$,
where $c(i) \subset \{1,\ldots,i-1\}$ is a conditioning index set of size $|c(i)| = \min(m,i-1)$ for all $i=2,\ldots,p$ (and $c(1) = \emptyset$). Next, we discuss the theoretical limitation of the proposed approximation for $\GP$ mixtures.

\noindent \textbf{Proposition 1\ [No Free Lunch Theorem].} Suppose $\adens ^{(\mbox{gpm, mle})}(\by)$ denotes the maximum likelihood estimator of $p^{(\mbox{gpm})}(\mathbf{y})$. Then,\\
$\mbox{KL}[p^{(\mbox{gpm})}(\mathbf{y}) \ ||\ \adens ^{(\mbox{gpm, mle})}(\by)]\leq $\\
$\min_{L_{1:G} \in \mathcal{S}_m} \bigg[\sum_{g=1}^G \pi_g\ \mbox{KL}[p(\mathbf{y}) \ ||\  \normal_p(\bfmu, (L_g L_g^{T}))^{-1}]\bigg]$\\
$:=\mbox{KL}_{\mbox{relaxed}}\bigg[p^{(\mbox{gpm})}(\mathbf{y}) \ ||\ \adens ^{(\mbox{gpm, vecchia(m)})}(\by)\bigg]$\\
for any choice of conditioning set size $m$.

\noindent\emph{Proof.} By the use of the definition of a maximum likelihood estimator under model misspecification \citep{5805f73c-4dfa-385e-bd6d-68424fb9f5be}, and the convexity of $\mbox{KL}$ divergence
, serially, we have\\
$\mbox{KL}[p^{(\mbox{gpm})}(\mathbf{y}) \ ||\ \adens ^{(\mbox{gpm, mle})}(\by)]$\\
$=\min_{L_{1:G} \in \mathcal{S}_m} \bigg[\mbox{KL}[p(\mathbf{y}) \ ||\ \sum_{g=1}^G \pi_g \normal_p(\bfmu, (L_g L_g^{T}))^{-1}]\bigg]$\\
$\leq\min_{L_{1:G}\in \mathcal{S}_m} \bigg[\sum_{g=1}^G \pi_g\ \mbox{KL}[p(\mathbf{y}) \ ||\  \normal_p(\bfmu, (L_g L_g^{T}))^{-1}]\bigg]$.
Hence, we have the proof.\\
Next, we discuss a theoretical result that provides crucial insights into the choice of the conditioning set size $m$.

\noindent \textbf{Proposition 2\ [Choice of Conditioning Set size].} For a choice of conditioning set size $m_1\geq m_2$, \\
$\mbox{KL}_{\mbox{relaxed}}\bigg[p^{(\mbox{gpm})}(\mathbf{y}) \ ||\ \adens ^{(\mbox{gpm, vecchia($m_1$)})}(\by)\bigg]$\\
$\leq \mbox{KL}_{\mbox{relaxed}}\bigg[p^{(\mbox{gpm})}(\mathbf{y}) \ ||\ \adens ^{(\mbox{gpm, vecchia($m_2$)})}(\by)\bigg]$.
That is, enlarging the conditioning sets  never increases the relaxed KL divergence.

\noindent\emph{Proof.} By definition, for $m_1\geq m_2$, we have $\mathcal{S}_{m_2}\subset \mathcal{S}_{m_1}$, which in turn implies\\ $\min_{L_{1:G} \in \mathcal{S}_{m_2}} \bigg[\sum_{g=1}^G \pi_g\ \mbox{KL}[p(\mathbf{y}) \ ||\  \normal_p(\bfmu, (L_g L_g^{T}))^{-1}]\bigg]$\\
$\geq \min_{L_{1:G} \in \mathcal{S}_{m_1}} \bigg[\sum_{g=1}^G \pi_g\ \mbox{KL}[p(\mathbf{y}) \ ||\  \normal_p(\bfmu, (L_g L_g^{T}))^{-1}]\bigg]$.
Hence, we have the proof.

\subsection{Vecchia-assisted EM (VEM) Algorithm \label{sec:vecchiaem}}
We  now have  the ground work in place to propose a computationally efficient modified EM algorithm that, first crucially exploits a systematic ordering \citep{Guinness2016a} of the points in the input grid of the $\GP$s, followed by a Vecchia approximation assisted fast Cholesky decomposition of large matrices \citep{Jurek2020}, to ensure scalability of clustering of $\GP$s observed over large grids. We first outline these data preprocessing steps below.
\vspace{-5pt}

\noindent\textbf{(i) Ordering.} First, following the popular choice in the literature, we order the points in the input grid via a maximin ordering \citep{Guinness2016a}. Then, we fix the maximum number of nearest neighbors in \eqref{eq:vecchia}, i.e, $|c(i)| \leq m \forall i = 1, \ldots, p$. We recall from Section \ref{sec: vecchiaintro} that, this induces a sparsity structure in the corresponding covariance matrices. The sparsity pattern, denoted by $\mathcal{S}$, remains fixed throughout the iterations of the algorithm and consequently results in substancial computational gain.
\vspace{-5pt}

\noindent\textbf{(ii) Matrix inversion.} We adopt a sparse Cholesky factorization  scheme for the covariance matrices $\bK_{g}$ imposing the sparsity structure $\mathcal{S}$ \citep{Jurek2020}, described in Algorithm \ref{alg:ic0}. Modulo $\mathcal{S}$, these sparse Cholesky approximations are calculated in $\order(pm^2)$ time.
\begin{algorithm}
\caption{Sparse Cholesky decomposition}
\vspace{-10pt}
\begin{itemize}[itemsep = -5pt]
    \item \textbf{Input} covariance matrix $\bfSigma \in \mathbb{R}^{p \times p}$, sparsity matrix $\mathcal{S} \in \{0,1\}^{p \times p}$
    \item \textbf{Output} lower-triangular $p\times p$ matrix $\bL = VeccInv(\bfSigma, \mathcal{S})$ 
\end{itemize}
\vspace{-8pt}
\begin{algorithmic}[1]
\FOR{$i=1$ to $p$}
\FOR{$j=1$ to $i-1$}
\IF{$\mathcal{S}_{i,j}=1$}
\STATE $\bL_{i,j} = (\bfSigma_{i,j} - \sum_{u=1}^{j-1}\bL_{i,u}\bL_{j,u})/\bL_{j,j}$ \\
\ENDIF
\ENDFOR 
\STATE $\bL_{i,i} = (\bfSigma_{i,i} - \sum_{u=1}^{i-1}\bL_{u,u})^{1/2}$
\ENDFOR
\end{algorithmic}
\label{alg:ic0}
\end{algorithm}

 We will be going to use this sparse Cholesky factorization algorithm throughout our EM algorithm, as stated below.

\noindent\textbf{(1) E-step.} From the \textbf{E} step Section \ref{sec:naiveem}, it is evident that the calculation of $W_{ig}$ 's require $\order(p^3)$ time for matrix computations due to inversion and determinant calculation of $\bK_{ig}$'s. Instead, we first calculate a sparse inverse Cholesky factor $\hat{\bU}_{g}$ = $VeccInv(\bK_{g}, \mathcal{S})$ using \ref{alg:ic0} in $\order(m\mbox{log} m)$ time. Consequently, we approximate $|\bK_{g}|$ by product of diagonal elements of $\bU_{g}$. The modified loglikelihood can be expressed as $\hat{Q}(\theta | \theta^{(s)}) = \sum_{i=1}^N \sum_{g=1}^G\left( \hat{W}^{(s)}_{ig} \mbox{log}(\pi_g \normal_p(\by_i; 0, \hat{K}_g))\right)$,
where $\hat{K}_g = [\hat{\bU}_{g}\hat{\bU}_{g}^\top]^{-1}$, and $\hat{W}^{(s)}_{ig}$'s denote the conditional expectations calculated using Vecchia approximation. .  

\noindent\textbf{(2) M-step.} Here, we modify Section \ref{sec:naiveem} as, $\pi^{(s+1)}_g \propto \sum_{i=1}^N \hat{W}^{(s)}_{ig}$, $l^{(s+1)}_g = l^{(s)}_g + \lambda\frac{d}{dl_g}\hat{\epsilon}^{(s)}$, $\sigma^{(s+1)}_g = \sigma^{(s)}_g + \lambda\frac{d}{d\sigma_g}\hat{\epsilon}^{(s)}$,
where $\hat{\epsilon}^{(s)} = \sum_{i,g} \frac{\hat{W}^{(s)}_{ig}}{2}\left(\mbox{log}|\hat{\bK}^{-1}_g| - \by^\top_i \hat{\bK}^{-1}_g \by_i\right)$.


\section{Performance Evaluation}\label{sec:performance}
\begin{figure}[htp]
    \centering
    \begin{subfigure}{0.22\textwidth}
    \centering
        \includegraphics[width = 1\textwidth]{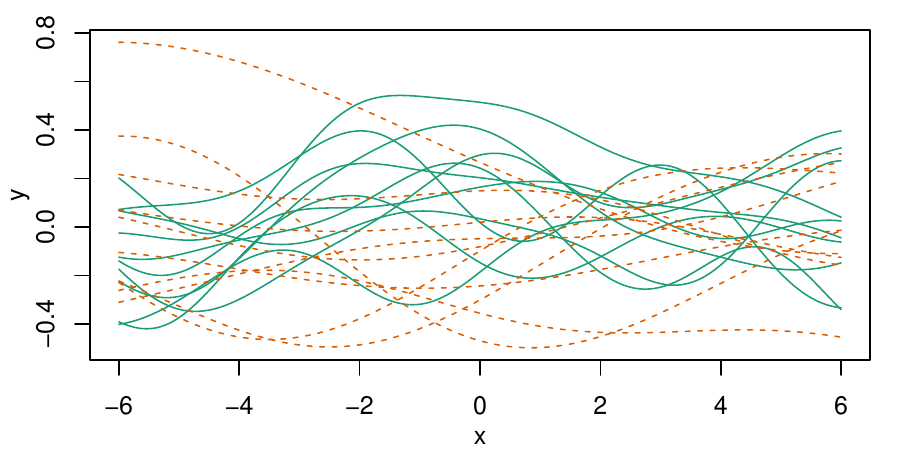}
    \caption{Simulated data (scenario 1)}
    \label{subfig:gpdata_case1}
    \end{subfigure}
    \hspace{1mm}
    \begin{subfigure}{0.22\textwidth}
    \centering
        \includegraphics[width = 1\textwidth]{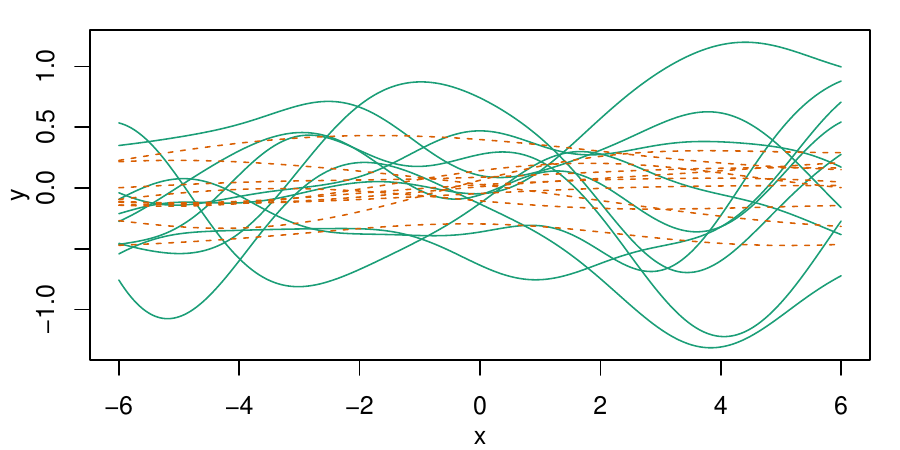}
    \caption{Simulated data (scenario 2)}
    \label{subfig:gpdata_case2}
    \end{subfigure}
    \caption{Simulated Gaussian process realizations}
    \label{fig:gpdata}
\end{figure}

\noindent\textbf{Experiments.}
From each of the two Gaussian processes $\by(\cdot) \sim \GP(0,\ \bK_{l_i, \sigma_i}),\ i = 1, 2$, we simulate $10$ realizations observed over an uniform grid of length $p$ = $300$; and consider the task of classifying the $20$ realizations into $2$ clusters. We compare the accuracy and speed of the naive EM algorithm (\ref{sec:naiveem}) and the proposed V
EM algorithm (\ref{sec:vecchiaem}) with varying values of conditioning set size $m\in\{15,25,30,60\}$. We utilise the Normalized Mutual Information (NMI) \citep{nmi} as a measure of agreement between the true and the estimated clusters. 
We consider $25$ repeated trials under two separate scenarios corresponding two sets of choices of the covariance kernel parameters: \textbf{(1)} a difficult case with $(l_1, \sigma_1) = (0.2, 0.2)$ and $(l_2, \sigma_2) = (0.5, 0.3)$, where two clusters are rather indistinguishable, as shown in \ref{subfig:gpdata_case1}; \textbf{(2)} an easy case with $(l_1, \sigma_1) = (0.2, 0.5)$ and $(l_2, \sigma_2) = (0.5, 0.2)$, where the clusters are distinguishable, as shown in figure \ref{subfig:gpdata_case2} . 

In the first scenario, the V
EM algorithm often performs similar to the  computationally intensive naive EM algorithm as $m$ increases. For lower the values of $m$, the Vecchia approximation of $\GP$ mixtures renders itself inaccurate and the clustering accuracy via the V
EM algorithm deteriorates significantly, as expected. Refer to Propositions 1 and 2 in Section \ref{sec:vecchiaem} for further discussions. 
\begin{figure}[!htbp]
    \centering
    \begin{subfigure}{0.22\textwidth}
    \centering
        \includegraphics[width = 1\textwidth]{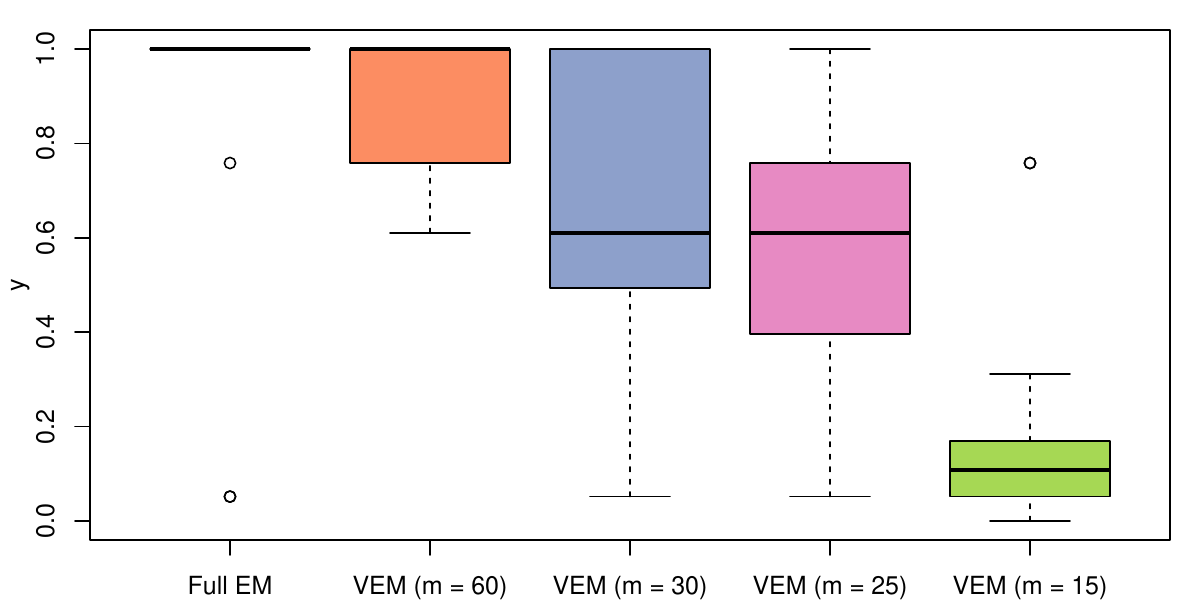}
    \caption{Scenario 1}
    \label{subfig:boxplot_case1}
    \end{subfigure}
    \hspace{1mm}
    \begin{subfigure}{0.22\textwidth}
    \centering
        \includegraphics[width = 1\textwidth]{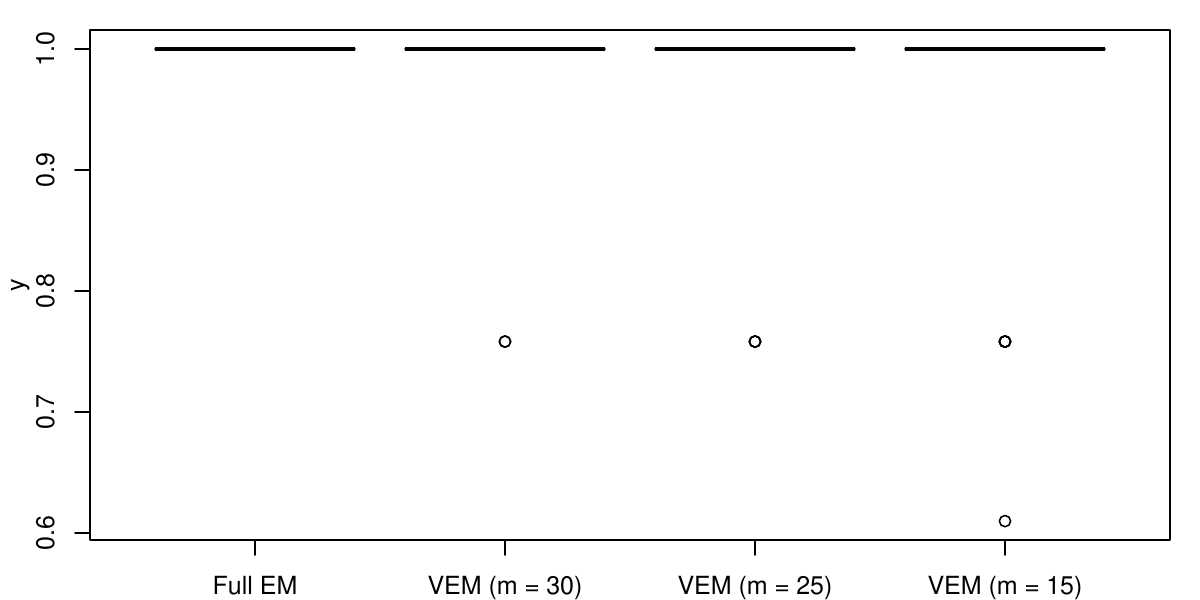}
    \caption{Scenario 2} 
    \label{subfig:boxplot_case2}
    \end{subfigure}
    \caption{Boxplot of NMI values for multiple simulations of the two scenarios in figure \ref{fig:gpdata}. Higher NMI values are desired.}
    \vspace{-5pt}
    \label{fig:boxplot}
\end{figure}
In the second scenario, the computationally efficient V
EM algorithm almost always performs similar to the 
naive EM algorithm even a very small $m$, while enjoying remarkably improved time complexity.
For instance, with $m /p = 0.1$ (i.e $m$ = 30), the VEM algorithm took on an average only 40$\%$ of the time taken by the naive EM algorithm. The computational advantage gets exponentially better for moderate to large data-set. For example, with $p =700$, VEM algorithm with the same $m/p$ ratio took only 10$\%$ time of that of the naive EM algorithm.

\begin{figure}[ht]
    \centering
    \includegraphics[width = 0.45\textwidth]{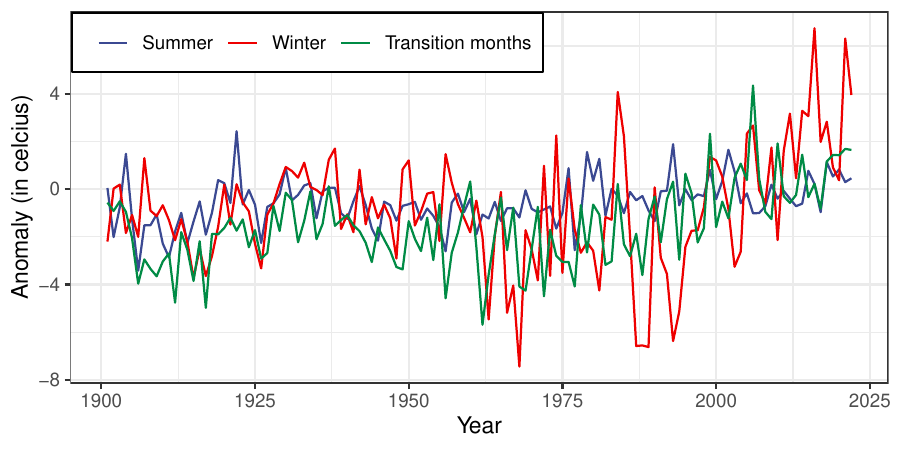}
    \caption{Estimated mean temperature anomalies in each cluster}
    \vspace{-5pt}
    \label{fig:tempanomalymeans}
\end{figure}
\noindent\textbf{Temperature Anomalies at the North Pole.}  We consider data on the monthly temperature anomalies in the Earth's geographic north pole between years $1901-2022$, publicly available at the National Oceanic and Atmospheric Administration website (\href{https://www.ncei.noaa.gov/access/monitoring/climate-at-a-glance/global/time-series}{noaa.gov}). Due to numerous anthropogenic factors, including rapid industrialization, agriculture, burning of fossil fuel 
 etc., temperature has increasingly fluctuated 
 since 1901. The study of the temperature anomaly in the Earth's north pole is especially critical, since, among other impacts,  the weakened polar jet stream often  brings the polar vortex further south, which results in extreme weather events in North America, Europe and Asia. 
The goal of this study is to group the $12$ months of year with respect to the extent of temperature anomaly, based on the perception that the weather is impacted differently in the winter and summer months at the north pole. To that end, we consider $12$ time series of length $122$, one corresponding to each month of the year, of  monthly temperature anomalies over the years $1901-2022$. We take $5$-year moving averages for each of the $12$ time series to account for the cyclical variations, and consider $\GP$'s with the Matern covariance kernel with $\nu = 0.5$ \citep{porcu2023matern} to describe the resulting time-series.  Considering $3$ classes, we carry out the clustering task via  the naive EM algorithm \ref{sec:naiveem} and the proposed V
EM algorithm \ref{sec:vecchiaem} with conditioning set size $m=10$. The result from computationally efficient V
EM algorithm matches with that obtained via the computationally intensive naive EM algorithm, and  it is displayed in Figure \ref{fig:tempanomalymeans}. Three distinct clusters are identified, e.g,  summer months (June to August), winter months (October to February), and transitioning months (March, April and September).

\section{Conclusion}
Taking advantage of  the popular Vecchia approximation of $\GP$s, we proposed an efficient EM algorithm for $\GP$ clustering, and delineated the computational gains. Theoretical limitations of the proposed methodology is presented via a no free lunch theorem. In future, we shall develop a  Vecchia assisted non-parametric Bayes  $\GP$ clustering algorithm, that simultaneously learns the numbers of the clusters and the clustering indices from the data.
\bibliographystyle{plainnat}
\bibliography{paper-ref.bib, additionalrefs.bib}

\end{document}